\documentclass[12pt,german,english]{article}
\usepackage[T1]{fontenc}
\usepackage[latin9]{inputenc}
\usepackage{geometry}
\usepackage{color}
\usepackage{amsmath}
\usepackage{amssymb}
\usepackage{setspace}
\makeatletter
\newcommand{\lyxaddress}[1]{
\par {\raggedright #1
\vspace{1.4em}
\noindent\par}
}

\usepackage[dvips]{graphicx}            
\usepackage{pslatex}

\usepackage[T1]{fontenc}
\usepackage{epsfig,url}
\usepackage[latin9]{inputenc}
\usepackage{geometry}
\usepackage{amsthm}
\usepackage{enumitem}
\usepackage{amstext}
\usepackage{setspace}

\usepackage{pstricks}
\usepackage{graphicx}
\usepackage{epsf}
\usepackage{epsfig}

\usepackage{babel}

\usepackage{lipsum}

\let\OLDthebibliography\thebibliography
\renewcommand\thebibliography[1]{
  \OLDthebibliography{#1}
  \setlength{\parskip}{0pt}
  \setlength{\itemsep}{0pt plus 0.3ex}
}
\usepackage{amssymb}

\usepackage[numbers,sort&compress]{natbib}

\makeatother

\usepackage{babel}
\begin{document}

\title{The Elliptic Sunrise \date{}}

\author{Luise Adams $^{a}$, Christian Bogner $^{b}$ and Stefan Weinzierl
$^{a}$}

\maketitle
\begin{center}
$^{a}$\emph{PRISMA Cluster of Excellence, Institut f\"ur Physik,
}\\
\emph{Johannes Gutenberg-Universit\"at Mainz,}\\
\emph{D - 55099 Mainz, Germany}
\par\end{center}

\lyxaddress{\begin{center}
\emph{$^{b}$Institut f\"ur Physik, Humboldt-Universit\"at zu Berlin,
}\\
\emph{D - 10099 Berlin, Germany}
\par\end{center}}
\begin{abstract}
In this talk, we discuss our recent computation of the two-loop sunrise
integral with arbitrary non-zero particle masses. In two space-time
dimensions, we arrive at a result in terms of elliptic dilogarithms.
Near four space-time dimensions, we obtain a result which furthermore
involves elliptic generalizations of Clausen and Glaisher functions.

\pagebreak{}
\end{abstract}

\section{Introduction}

In the computation of many Feynman integrals the use of multiple polylogarithms
\cite{Gon} 
\[
\textrm{Li}_{(s_{1},\,...,\, s_{k})}(z_{1},\,...,\, z_{k})=\sum_{n_{1}>n_{2}>...>n_{k}\geq1}\frac{z_{1}^{n_{1}}...z_{k}^{n_{k}}}{n_{1}^{s_{1}}...n_{k}^{s_{k}}},\, s_{i}\geq1,\,|z_{i}|<1
\]
is very advantageous. In particular, these functions, shown as nested
sums here, also have representations as iterated integrals, given
by the classes of hyperlogarithms \cite{Lap1,Lap2} or by iterated
integrals on moduli spaces of curves of genus zero (see \cite{Bro}).
Apparently, it is not possible to express every Feynman integral in
terms of this framework of functions. This problem is expected to
affect an entire class of massive integrals (see e.g. \cite{Bauetal})
and was furthermore pointed out for certain massless integrals, arising
in $\mathcal{N}$ = 4 supersymmetric Yang-Mills theory \cite{CarLar,Nanetal}. 

One of the simplest Feynman integrals where multiple polylogarithms
are not sufficient to express the result is the massive two-loop sunrise
integral
\[
S(D,\, t)=\int\frac{d^{D}k_{1}d^{D}k_{2}}{\left(i\pi^{D/2}\right)^{2}}\frac{1}{\left(-k_{1}^{2}+m_{1}^{2}\right)\left(-k_{2}^{2}+m_{2}^{2}\right)\left(-\left(p-k_{1}-k_{2}\right)^{2}+m_{3}^{2}\right)}.
\]
In this talk, we consider this integral as a function of the three
particle masses satisfying $0<m_{1}\leq m_{2}\leq m_{3}<m_{1}+m_{2}$
and of the squared momentum $t=p^{2}.$ We omit an explicit mass-scale
parameter $\mu$ in our equations. We discuss the computation of this
Feynman integral at $D=2$ and $D=4$ dimensions in terms of the Laurent
expansions
\begin{eqnarray*}
S(2-2\epsilon,\, t) & = & {\color{black}{\color{red}{\color{black}S^{(0)}(2,\, t)}}+{\color{red}{\color{blue}{\color{black}S^{(1)}(2,\, t)}}}}\epsilon+\mathcal{O}\left(\epsilon^{2}\right),\\
S(4-2\epsilon,\, t) & = & S^{(-2)}(4,\, t)\epsilon^{-2}+S^{(-1)}(4,\, t)\epsilon^{-1}+{\color{red}{\color{blue}{\color{black}S^{(0)}(4,\, t)}}}+\mathcal{O}(\epsilon).
\end{eqnarray*}
In the case of $D=2$, the integral is finite and our result is the
coefficient ${\color{black}S^{(0)}(2,\, t)}$. In the case of $D=4,$
we compute the coefficient ${\color{red}{\color{blue}{\color{black}S^{(0)}(4,\, t)}}}.$
The pole terms were already known and read
\begin{eqnarray*}
S^{(-2)}(4,\, t) & = & -\frac{1}{2}\left(m_{1}^{2}+m_{2}^{2}+m_{3}^{2}\right),\\
S^{(-1)}(4,\, t) & = & \frac{1}{4}t-\frac{3}{2}\left(m_{1}^{2}+m_{2}^{2}+m_{3}^{2}\right)+\sum_{i=1}^{3}m_{i}^{2}\ln\left(m_{i}^{2}\right).
\end{eqnarray*}
 In order to obtain ${\color{blue}{\color{black}S^{(0)}(4,\, t)}},$
we compute the $\epsilon$-coefficient $S^{(1)}(2,\, t)$ of the two-dimensional
case and relate $S(2-2\epsilon,\, t)$ with $S(4-2\epsilon,\, t)$
by Tarasov's dimension shift relations \cite{Tar1,Tar2}. Our work
on these integrals is motivated by the search for classes of functions
beyond multiple polylogarithms, which are appropriate for the computation
of Feynman integrals. 

In section \ref{sec:Basic-properties-of} we briefly comment on three
computational approaches which fail to provide a result in terms of
multiple polylogarithms for the massive sunrise integral. We begin
our computation with the integral in two dimensions and discuss our
first solution of the differential equation for ${\color{black}S^{(0)}(2,\, t)}$
in section \ref{sec:A-solution-to}. In section \ref{sec:An-elliptic-dilogarithm}
we express this result in terms of an elliptic dilogarithm. Section
\ref{sec:Elliptic-Clausen--and} introduces further elliptic generalizations
of polylogarithms, understood as elliptic generalizations of Clausen
and Glaisher functions, which arise in our results for $S^{(1)}(2,\, t)$
and $S^{(0)}(4,\, t)$. Section \ref{sec:Conclusions} contains the
conclusions of this talk.

\section{Basic properties of the massive sunrise integral\label{sec:Basic-properties-of}}

The massive sunrise integral was extensively studied in the past \cite{BroFleTar,Bauetal,Bauetal2,Bauetal3,Beretal,BloVan,CafCzyRem,Caffoetal,CafGunRem,GroKoePiv,GroKoePiv2,LapRem,MueWeiZay,PozRem,RemTan,Baileyetal,KalKni,Smi,DavDel,Broadhu}.
Let us recall some important aspects. 

Firstly, in \cite{Beretal} the integral $S(D,\, t)$ is expressed
as a linear combination of generalized hypergeometric functions of
Lauricella type C, which are functions of $t$, of the squared particle
masses and of the dimension $D.$ While a wide range of generalized
hypergeometric functions can be expanded in terms of multiple polylogarithms
with today's methods, this has not been achieved for the mentioned
result so far. 

Secondly, one may attempt to compute the integral by integration over
Feynman parameters. In terms of Feynman parameters, the integral in
$D=2$ dimensions reads
\[
S(2,\, t)=\int_{\sigma}\frac{\omega}{\mathcal{F}},
\]
with $\omega=x_{1}dx_{2}\wedge dx_{3}+x_{2}dx_{3}\wedge dx_{1}+x_{3}dx_{1}\wedge dx_{2}$
and $\sigma=\left\{ \left[x_{1}:x_{2}:x_{3}\right]\in\mathbb{P}^{2}|x_{i}\geq0,\, i=1,\,2,\,3\right\} $
while the second Symanzik polynomial is given as 
\[
\mathcal{F}=-x_{1}x_{2}x_{3}t+\left(x_{1}m_{1}^{2}+x_{2}m_{2}^{2}+x_{3}m_{3}^{2}\right)\left(x_{1}x_{2}+x_{2}x_{3}+x_{1}x_{3}\right).
\]
For an attempt to iteratively build up the result in terms of the
mentioned iterated integrals which represent the multiple polylogarithms,
the polynomial $\mathcal{F}$ would have to satisfy the criterion
of linear reducibility \cite{Bro2}. The latter is a sufficient but
not necessary criterion to obtain multiple polylogarithms in the result.
However, the polynomial fails this criterion and a change of variables
to restore linear reducibility for a new set of integration variables
is unknown for this case. 

Thirdly, the integral \foreignlanguage{german}{$S(D,\, t)$ for generic
space-time dimension satisfies an inhomogeneous fourth-order differential
equation in $t$:
\begin{equation}
\left(P_{4}\frac{d^{4}}{dt^{4}}+P_{3}\frac{d^{3}}{dt^{3}}+P_{2}\frac{d^{2}}{dt^{2}}+P_{1}\frac{d^{1}}{dt^{1}}+P_{0}\right)S\left(D,\, t\right)=c_{12}T_{12}+c_{13}T_{13}+c_{23}T_{23}\label{eq:master dgl}
\end{equation}
where the $T_{ij}=T\left(m_{i}^{2},D\right)T\left(m_{j}^{2},D\right)$
are products of tadpole integrals 
\[
T(m^{2},D)=\int\frac{d^{D}k}{i\pi^{\frac{D}{2}}}\frac{1}{\left(-k^{2}+m^{2}\right)}=\Gamma\left(1-\frac{D}{2}\right)\left(m^{2}\right)^{\frac{D}{2}-1}.
\]
 All coefficients $P_{k}$ and $c_{ij}$ are polynomials in $m_{1}^{2},\, m_{2}^{2},\, m_{3}^{2},\, t,\, D.$}
\foreignlanguage{german}{Each of the functions $S^{(0)}(2,\, t),$}
$S^{(1)}(2,\, t),$ $S^{(0)}(4,\, t)$\foreignlanguage{german}{ satisfies
an inhomogeneous differential equation of second or higher order.}
If any of these operators would factorize into differential operators
of first order the corresponding coefficient could be obtained as
an iterated integral in a straightforward way (see e.g. section 2
of \cite{AdaBogWei1}). However, this is not the case for any of these
operators. 

All of these points give rise to the expectation, that we need functions
beyond multiple polylogarithms to express the integrals \foreignlanguage{german}{$S^{(0)}(2,\, t),$}
$S^{(1)}(2,\, t),$ $S^{(0)}(4,\, t)$.\foreignlanguage{german}{ This
expectation is confirmed by our results for these functions.}

\section{The differential equation in two dimensions \label{sec:A-solution-to}}

We follow the approach of differential equations and begin with the
Feynman integral in $D=2$ dimensions. For the case of equal masses
$m_{1}=m_{2}=m_{3},$ a differential equation of second order was
already given in \cite{BroFleTar}. A full solution in terms of integrals
over elliptic integrals was obtained in \cite{LapRem}.

For the case of arbitrary masses, a differential equation of second
order was found later in \cite{MueWeiZay}:

\selectlanguage{german}%
\begin{eqnarray}
L_{2}\, S(2,\, t) & = & p_{3}(t),\nonumber \\
L_{2} & = & p_{2}(t)\frac{d^{2}}{dt^{2}}+p_{1}(t)\frac{d}{dt}+p_{0}(t),\label{eq:L2 operator}
\end{eqnarray}
where $p_{0}(t),\, p_{1}(t),\, p_{2}(t)$ are polynomials in $t$
and in the $m_{i}^{2}$ and where $p_{3}(t)$ furthermore involves
$\ln(m_{i}^{2}),\, i=1,\,2,\,3.$ We take this equation as the starting
point of our computation and make the classical ansatz 
\begin{equation}
S(2,\, t)=C_{1}\psi_{1}(t)+C_{2}\psi_{2}(t)+\int_{0}^{t}dt_{1}\frac{p_{3}(t_{1})}{p_{0}(t_{1})W(t_{1})}\left(-\psi_{1}(t)\psi_{2}(t_{1})+\psi_{2}(t)\psi_{1}(t_{1})\right)\label{eq:diff eq dim 2}
\end{equation}
where $\psi_{1},$ $\psi_{2}$ are solutions of the homogeneous equation,
$C_{1},\, C_{2}$ are constants and 
\[
W(t)=\psi_{1}(t)\frac{d}{dt}\psi_{2}(t)-\psi_{2}(t)\frac{d}{dt}\psi_{1}(t)
\]
 is the Wronski determinant.

At this point, it is useful to consider the zero-set of the second
Symanzik polynomial $\mathcal{F}.$ This cubical curve intersects
the integration domain $\sigma$ of the Feynman integral at the three
points \foreignlanguage{english}{
\[
P_{1}=[1:0:0],\; P_{2}=[0:1:0],\; P_{3}=[0:0:1].
\]
 }We choose one of these points $P_{i}$ as the origin and transform
the curve to Weierstrass normal form 
\begin{equation}
y^{2}z-x^{3}-g_{2}(t)xz^{2}-g_{3}(t)z^{3}=0.\label{eq:Weierstrass}
\end{equation}
By this transformation, the chosen origin is mapped to the point $[x:y:z]=[0:1:0]$.
In this way, we obtain three elliptic curves $E_{\mathcal{F},i}$
according to the three points $P_{i},$ $i=1,\,2,\,3.$

In the chart $z=1$ we write eq. \ref{eq:Weierstrass} as 
\[
y^{2}=4(x-e_{1})(x-e_{2})(x-e_{3}),
\]
which defines the three roots $e_{1},\, e_{2},\, e_{3}$ with $e_{1}+e_{2}+e_{3}=0.$
These provide the boundaries of the period integrals 
\[
\psi_{1}=2\int_{e_{2}}^{e_{3}}\frac{dx}{y}=\frac{4}{\tilde{D}^{\frac{1}{4}}}K(k),\;\,\psi_{2}=2\int_{e_{1}}^{e_{3}}\frac{dx}{y}=\frac{4i}{\tilde{D}^{\frac{1}{4}}}K(k')
\]
of the elliptic curve. Here the polynomial $\tilde{D}$ is given as
\[
\tilde{D}=(t-(m_{1}+m_{2}-m_{3})^{2})(t-(m_{1}-m_{2}+m_{3})^{2})(t-(-m_{1}+m_{2}+m_{3})^{2})(t-(m_{1}+m_{2}+m_{3})^{2})
\]
and we have obtained the complete elliptic integral of the first kind
\[
K(x)=\int_{0}^{1}dt\frac{1}{\sqrt{(1-t^{2})(1-x^{2}t^{2})}}
\]
with moduli $k=\sqrt{\frac{e_{3}-e_{2}}{e_{1}-e_{2}}},$ $k'=\sqrt{1-k^{2}}=\sqrt{\frac{e_{1}-e_{3}}{e_{1}-e_{3}}}.$
These period integrals $\psi_{1},\,\psi_{2}$ are solutions of the
homogeneous equation associated to eq. \ref{eq:diff eq dim 2}.

We still have to fix the constants. It can be shown that $C_{2}$
has to vanish while the other constant $C_{1}$ is derived from a
known result \cite{UssDav,Bernetal,LuPer} for the zero-mass limit
$S(2,\,0).$ Now all pieces of our ansatz in eq. \ref{eq:diff eq dim 2}
are determined. In order to simplify the integrand of the particular
solution, we furthermore make use of the remaining two associated
period integrals of $E_{\mathcal{F},i}.$ In conclusion, we obtain
a result \cite{AdaBogWei1} of the form 
\begin{equation}
S(2,\, t)=S(2,\,0)+\frac{\psi_{1}(t)}{\pi^{2}}\int_{0}^{t}dt_{1}\rho(t_{1})\label{eq:result 2-dim}
\end{equation}
where the integrand $\rho$ involves elliptic integrals of the first
and second kind. 

\selectlanguage{english}%

\section{The massive sunrise integral in two dimensions\label{sec:An-elliptic-dilogarithm}}

\selectlanguage{german}%
The general shape of our result of eq. \ref{eq:result 2-dim} has
a disadvantage. While the involved elliptic integrals are well-studied
functions, nicely related to the underlying elliptic curve of the
problem, the integral over these functions in not a known function.
This integral might remind us vaguely of an iterated integral, but
in this form, it can not be recognized as a generalization of a polylogarithm.
However, for the equal-mass case, it was shown more recently in \cite{BloVan},
that the integral can be expressed in terms of an elliptic dilogarithm.
Various notions of elliptic polylogarithms were previosly introduced
in the mathematical literature \cite{BeiLev,Blo,BroLev,Lev,LevRac,Wil}. 

\selectlanguage{english}%
Before we apply an elliptic generalization of a polylogarithm to the
sunrise integral with arbitrary masses, let us briefly recall the
basic concept of an elliptic function. \foreignlanguage{german}{With
respect to a lattice $L=\mathbb{Z}+\tau\mathbb{Z}$ with $\tau\in\mathbb{C}$
and Im$(\tau)>0,$ a function $f$ is said to be elliptic, if it satisfies
$f(x)=f(x+\lambda)$ for $\lambda\in L.$ Accordingly, the corresponding
function $\tilde{f}(z)$ of $z\in\mathbb{C}^{\star}$ defined by $\tilde{f}(e^{2\pi ix})=f(x)$
is elliptic, if 
\begin{equation}
\tilde{f}\left(z\right)=\tilde{f}\left(z\cdot q_{\lambda}\right),\; q_{\lambda}\in e^{2\pi i\lambda}\textrm{ for }\lambda\in L.\label{eq:ellipticity}
\end{equation}
Recall that a cell of the lattice with $\tau=\frac{\psi_{2}}{\psi_{1}}$
is isomorphic to an elliptic curve with the periods $\psi_{1},\,\psi_{2}.$ }

\selectlanguage{german}%
A crucial idea for the construction of such elliptic functions is
to consider sums of the form $\sum_{n\in\mathbb{Z}}g\left(z\cdot q^{n}\right)$
over some function $g.$ If a sum of this type is well-defined, it
clearly satisfies the condition of eq. \ref{eq:ellipticity} by construction.
This concept can serve for definitions of elliptic generalizations
of polylogarithms. For example in \cite{BroLev} it is used to define
the class of multiple elliptic polylogarithms. The elliptic dilogarithm
in this framework reads
\[
\tilde{E}_{2}(z;\, u;\, q)=\sum_{m\in\mathbb{Z}}u^{m}\textrm{Li}_{2}(q^{m}z)
\]
where $u$ is a sufficiently small damping parameter to guarantee
the convergence of the function.

Based on the same basic idea, we define the class of functions \cite{AdaBogWei2}
\[
\textrm{ELi}_{n;m}(x;y;q)=\sum_{j=1}^{\infty}\sum_{k=1}^{\infty}\frac{x^{j}}{j^{n}}\frac{y^{k}}{k^{m}}q^{jk}=\sum_{k=1}^{\infty}\frac{y^{k}}{k^{m}}\textrm{Li}_{n}(q^{k}x),
\]

\begin{equation}
\textrm{E}_{n;\, m}(x;\, y;\, q)=\begin{cases}
\frac{1}{i}\left(\frac{1}{2}\textrm{Li}_{n}(x)-\frac{1}{2}\textrm{Li}_{n}(x^{-1})+\textrm{ELi}_{n;\, m}(x;\, y;\, q)-\textrm{ELi}_{n;\, m}(x^{-1};\, y^{-1};\, q)\right) & \textrm{ for }n+m\textrm{ even,}\\
\frac{1}{2}\textrm{Li}_{n}(x)+\frac{1}{2}\textrm{Li}_{n}(x^{-1})+\textrm{ELi}_{n;\, m}(x;\, y;\, q)+\textrm{ELi}_{n;\, m}(x^{-1};\, y^{-1};\, q) & \textrm{ for }n+m\textrm{ odd.}
\end{cases}\label{eq:Definition ell polylog E}
\end{equation}
Note that our elliptic dilogarithm 
\[
\textrm{E}_{2;\,0}(x;\, y;\, q)=\frac{1}{i}\left(\frac{1}{2}\textrm{Li}_{2}\left(x\right)-\frac{1}{2}\textrm{Li}_{2}\left(x^{-1}\right)+\sum_{i=1}^{\infty}y^{i}\textrm{Li}_{2}\left(q^{i}x\right)-\sum_{j=1}^{\infty}y^{-j}\textrm{Li}_{2}\left(q^{j}x^{-1}\right)\right)
\]
 is closely related to the above function $\tilde{E}_{2}.$ We obtain
\begin{eqnarray*}
\textrm{E}_{2;\,0}(x;\, y;\, q) & = & \frac{1}{i}\left(\tilde{E}_{2}(x;\, y;\, q)-\frac{1}{2}\frac{1+y}{1-y}\zeta(2)-\frac{1}{4}\frac{1+y}{1-y}\ln^{2}(-x)\right.\\
 &  & \left.-\frac{y}{(1-y)^{2}}\ln(-x)\ln(q)-\frac{1}{2}\frac{y\left(1+y\right)}{(1-y)^{3}}\ln^{2}(q)\right)
\end{eqnarray*}
in the region of parameters given by $x\in\mathbb{C}\backslash[0,\,\infty[,$
$|y|>1$ and real-valued $q$ in the range $0\leq q<\min\left(|x|,\,\frac{1}{|x|},\,|y|,\,\frac{1}{|y|}\right).$ 

Using the function $\textrm{E}_{2;\,0},$ we express our result for
the massive sunrise integral in two space-time dimensions in a very
compact way as%
\footnote{By a slight abuse of notation, we denote with $\psi_{1}$ the above
function of $t$ and the corresponding function of $q$.%
}
\begin{equation}
S\left(2,\, t\right)=\frac{\psi_{1}(q)}{\pi}\sum_{i=1}^{3}\textrm{E}_{2;\,0}(w_{i}(q);\,-1;\,-q)\textrm{ where }q=e^{\pi i\frac{\psi_{2}(t)}{\psi_{1}(t)}}.\label{eq:result ellipt dilog}
\end{equation}
Note that the dependence on $t$ is now implicitly expressed in terms
of $q,$ which is defined by the periods of the elliptic curve. The
arguments $w_{1},\, w_{2},\, w_{3}$ are functions of $q$ and of
the squared particle masses. They are directly obtained from the three
intersection points $P_{1},\, P_{2},\, P_{3}$ by the consecutive
transformations on the elliptic curves $E_{\mathcal{F},i}$, $i=1,\,2,\,3,$
indicated above. In this sense, every piece of the compact result
eq. \ref{eq:result ellipt dilog} is nicely related to the underlying
elliptic curves $E_{\mathcal{F},i}$.

In the case of equal masses, the result simplifies to 
\[
S\left(2,\, t\right)=3\frac{\psi_{1}(q)}{\pi}\textrm{E}_{2;\,0}\left(\textrm{exp}\left(2\pi i/3\right);\,-1;\,-q\right).
\]

\selectlanguage{english}%

\section{The massive sunrise integral around four dimensions \label{sec:Elliptic-Clausen--and}}

By use of dimension shift relations \cite{Tar1,Tar2}, we express
the coefficient $S^{(0)}(4,\, t)$ of the sunrise integral near $D=4$
dimensions in terms of coefficients of the $D=2$ case \cite{AdaBogWei3}.
We obtain $S^{(0)}(4,\, t)$ as a linear combination of terms ${\color{red}{\color{black}S^{(0)}(2,\, t)}},\frac{\partial}{\partial m_{i}^{2}}{\color{red}{\color{black}S^{(0)}(2,\, t)}},\, S^{(1)}(2,\, t),\,\frac{\partial}{\partial m_{i}^{2}}S^{(1)}(2,\, t),\, i=1,\,2,\,3.$
Therefore, our remaining task is the computation of $S^{(1)}(2,\, t).$

From eq. \ref{eq:master dgl} we obtain the differential equation
\begin{equation}
L_{1,a}\, L_{1,b}\, L_{2}\, S^{(1)}(2,\, t)=I_{1}(t).\label{eq:diff eq for S1}
\end{equation}
Here $L_{1,a}$ and $L_{1,b}$ are differential operators of first
order, 
\[
L_{1,a}=p_{1,a}\frac{d}{dt}+p_{0,a}\textrm{ and }L_{1,b}=p_{1,b}\frac{d}{dt}+p_{0,b},
\]
where $p_{0,a},\, p_{1,a}$ are rational functions of $t$ and the
squared particle masses and $p_{0,b},\, p_{1,b}$ are polynomials
in these variables. The homogeneous solutions $\psi_{a},\,\psi_{b}$
of these operators, defined by 
\[
L_{1,a}\,\psi_{a}(t)=0\textrm{ and }L_{1,b}\,\psi_{b}(t)=0
\]
are easily obtained. 

The operator $L_{2}$ in eq. \ref{eq:diff eq for S1} is the one of
eq. \ref{eq:L2 operator} which already appeared in the differential
equation of the two-dimensional case. The inhomogeneous term $I_{1}$
of eq. \ref{eq:diff eq for S1} is a combination of certain differentiations
of our result ${\color{red}{\color{black}S^{(0)}(2,\, t)}}$, of logarithms
in the squared particle masses and of a polynomial in the squared
masses and in $t.$ 

Solving eq. \ref{eq:diff eq for S1} for the combination $L_{2}\, S^{(1)}(2,\, t),$
we obtain 
\begin{equation}
L_{2}\, S^{(1)}(2,\, t)=I_{2}(t)\label{eq:diff eq S1 mit L2}
\end{equation}
with 
\[
I_{2}(t)=\tilde{C}_{1}\psi_{b}(t)+\tilde{C}_{2}\psi_{b}(t)\int_{0}^{t}\frac{\psi_{a}(t_{1})dt_{1}}{p_{1,b}(t_{1})\psi_{b}(t_{1})}+\psi_{b}(t)\int_{0}^{t}\frac{\psi_{a}(t_{1})dt_{1}}{p_{1,b}(t_{1})\psi_{b}(t_{1})}\int_{0}^{t_{1}}\frac{I_{1}(t_{2})dt_{2}}{p_{1,a}(t_{2})\psi_{a}(t_{2})}
\]
where $\tilde{C}_{1},\,\tilde{C}_{2}$ are integration constants. 

Now with eq. \ref{eq:diff eq S1 mit L2} we have to solve a similar
differential equation as in the two-dimensional case, with the only
difference that the inhomogeneous part is more complicated. However,
we can make a similar ansatz and we have the same period integrals
$\psi_{1},\,\psi_{2}$ of $E_{\mathcal{F},i}$ as solutions of the
homogeneous equation. Therefore, it is useful to introduce the variable
$q$ again in the same way as in eq. \ref{eq:result ellipt dilog}.
In terms of integrals over $q,$ we obtain

\[
S^{(1)}(2,\, t)=C_{3}\psi_{1}+C_{4}\psi_{2}-\frac{\psi_{1}}{\pi}\int_{0}^{q}\frac{dq_{1}}{q_{1}}\int_{0}^{q_{1}}\frac{dq_{2}}{q_{2}}\frac{I_{2}\left(q_{2}\right)\psi_{1}\left(q_{2}\right)^{3}}{\pi p_{2}\left(q_{2}\right)W\left(q_{2}\right)^{2}}.
\]
The integration constants $C_{3},\, C_{4}$ are determined from boundary
conditions. Expanding the integrand, we can perform the integrations
order by order and obtain a $q-$expansion of $S^{(1)}(2,\, t)$ to
high orders. This step finally allows us to find a result for $S^{(1)}(2,\, t)$
in closed form, which can be confirmed to satisfy the differential
equation \ref{eq:diff eq S1 mit L2}. 

Let us refer to \cite{AdaBogWei3} for the explicit result and just
highlight some of its properties here. Apart from classical (multiple)
polylogarithms, the result involves the functions $\textrm{E}_{1;\,0}(x;\, y;\, q),$
$\textrm{E}_{2;\,0}(x;\, y;\, q),$ $\textrm{E}_{3;\,1}(x;\, y;\, q)$
as defined in eq. \ref{eq:Definition ell polylog E} and furthermore
a quadruple sum of the form 
\[
\Lambda\left(x_{1},x_{2};y_{1},y_{2};-q\right)=\sum_{j_{1}=1}^{\infty}\sum_{k_{1}=1}^{\infty}\sum_{j_{2}=1}^{\infty}\sum_{k_{2}=1}^{\infty}\frac{k_{1}^{2}\left(-q\right)^{j_{1}k_{1}+j_{2}k_{2}}}{j_{2}\left(j_{1}k_{1}+j_{2}k_{2}\right)^{2}}\left(x_{1}^{j_{1}}y_{1}^{k_{1}}-x_{1}^{-j_{1}}y_{1}^{-k_{1}}\right)\left(x_{2}^{j_{2}}y_{2}^{k_{2}}+x_{2}^{-j_{2}}y_{2}^{-k_{2}}\right).
\]
For the arguments of these functions, we have $y,\, y_{1},\, y_{2}\in\left\{ -1,\,1\right\} $
and $x,\, x_{1},\, x_{2}\in\left\{ w_{1},\, w_{2},\, w_{3}\right\} ,$
where the $w_{i}$ again are the arguments obtained from the intersection
points mentioned above. 

The appearance of the functions $\textrm{E}_{1;\,0}(x;\, y;\, q),\,\textrm{E}_{2;\,0}(x;\, y;\, q),\,\textrm{E}_{3;\,1}(x;\, y;\, q)$
shows that the framework of eq. \ref{eq:Definition ell polylog E},
set up for the coefficient $S^{(0)}(2,\, t)$, is also useful for
$S^{(1)}(2,\, t)$ and hence also for the four-dimensional case. Furthermore,
these functions can be viewed as elliptic generalizations of Clausen
and Glaisher functions. Recall that the Clausen functions are defined
by 
\[
\textrm{Cl}_{n}\left(\varphi\right)=\begin{cases}
\frac{1}{2i}\left(\textrm{Li}_{n}\left(e^{i\varphi}\right)-\textrm{Li}_{n}\left(e^{-i\varphi}\right)\right) & \textrm{ for even }n,\\
\frac{1}{2}\left(\textrm{Li}_{n}\left(e^{i\varphi}\right)+\textrm{Li}_{n}\left(e^{-i\varphi}\right)\right) & \textrm{ for odd }n,
\end{cases}
\]
and the Glaisher functions are given as
\[
\textrm{Gl}_{n}\left(\varphi\right)=\begin{cases}
\frac{1}{2}\left(\textrm{Li}_{n}\left(e^{i\varphi}\right)+\textrm{Li}_{n}\left(e^{-i\varphi}\right)\right) & \textrm{ for even }n,\\
\frac{1}{2i}\left(\textrm{Li}_{n}\left(e^{i\varphi}\right)-\textrm{Li}_{n}\left(e^{-i\varphi}\right)\right) & \textrm{ for odd }n.
\end{cases}
\]
We therefore obtain as 'non-elliptic limits' of our functions: 
\begin{eqnarray*}
\textrm{lim}_{q\rightarrow0}\textrm{E}_{1;\,0}\left(e^{i\varphi};\, y;\, q\right) & = & \textrm{Cl}_{1}\left(\varphi\right),\\
\textrm{lim}_{q\rightarrow0}\textrm{E}_{2;\,0}\left(e^{i\varphi};\, y;\, q\right) & = & \textrm{Cl}_{2}\left(\varphi\right),\\
\textrm{lim}_{q\rightarrow0}\textrm{E}_{3;\,1}\left(e^{i\varphi};\, y;\, q\right) & = & \textrm{Gl}_{3}\left(\varphi\right).
\end{eqnarray*}
As a final remark, let us mention that $S^{(1)}(2,\, t)$ is a function
of mixed weight. It shares this property with the function $\textrm{E}_{3;\,1}(x;\, y;\, q)$
which has parts of weight three and of weight four.

\section{Conclusions\label{sec:Conclusions}}

We discussed the computation of the massive sunrise integral in two
and around four space-time dimensions. We started with the computation
of the $\mathcal{O}\left(\epsilon^{0}\right)$-part of the integral
in two dimensions and expressed our result in terms of an elliptic
dilogarithm. In this form, the result is very compact and every part
of it is nicely related to the underlying elliptic curve, given by
the second Symanzik polynomial of the Feynman graph. 

We continued with the computation of the $\mathcal{O}\left(\epsilon^{1}\right)$-part
in two dimensions. Apart from the elliptic dilogarithm, this result
involves further elliptic generalizations of (multiple) polylogarithms,
which can be understood as elliptic generalizations of Clausen and
Glaisher functions. Due to well-known dimension shift relations, these
results provide the $\mathcal{O}\left(\epsilon^{0}\right)$-part of
the Feynman integral in four dimensions. 

Together with the results of \cite{BloVan,BloKerVan}, our results
give rise to the hope, that elliptic (multiple) polylogarithms may
serve as an appropriate class of functions to compute further Feynman
integrals beyond multiple polylogarithms. Some of our functions can
be related to the functions of \cite{BroLev}, where also a framework
of iterated integrals, already applied in a different physics context
\cite{BroMafMatSch}, is provided.

\end{document}